%% file: apssamp.tex
\documentclass[%
reprint,
 amsmath,amssymb,
 aps,prl,
]{revtex4-2}

\usepackage{hyperref}
\usepackage{cleveref}
\input{cleveref_setup}

\input{macros}

\renewcommand{\rhosound}{{\rho_s}}

\renewcommand{\LTi}{L_{Ti}}

\newcommand{\omegac}{\omega_c}
\newcommand{\lmin}{\ell_\text{min}}
\newcommand{\lmax}{\ell_\text{max}}

\usepackage{graphicx}%
\usepackage{dcolumn}%
\usepackage{bm}%
\usepackage{xcolor}

\usepackage{tikz}
\usetikzlibrary{calc}

\begin{document}

\crefname{section}{Sec.}{Secs.}
\Crefname{section}{Section}{Sections}

\crefname{appendix}{Appendix}{Appendices}
\Crefname{appendix}{Appendix}{Appendices}

\title{Destabilization of temperature-gradient-driven plasma turbulence by equilibrium \exb{} flow shear}

\author{Haomin Sun$^{1,*,\ddagger}${, }Plamen G. Ivanov$^{2,\dagger,\ddagger}$, Justin Ball$^{1}$, Stephan Brunner$^{1}$, {and }Bhavin S. Patel$^{2,3}$}
\affiliation{$^1$Ecole Polytechnique F\'ed\'erale de Lausanne (EPFL), Swiss Plasma Center (SPC), CH-1015 Lausanne, Switzerland\\
$^2$UKAEA (United Kingdom Atomic Energy Authority), Culham Campus, Abingdon, Oxfordshire, OX14 3DB, United Kingdom of Great Britain and Northern Ireland\\
$^3$Gridfire Inc., Wilmington, Delaware 19808, USA\\
$*$Corresponding author: haomin.sun@epfl.ch\\
$\dagger$Corresponding author: plamen.ivanov@ukaea.uk\\
$\ddagger$ These two authors {contributed} equally to this work.
}

\vspace{10pt}

\begin{abstract}

A novel physical mechanism whereby sheared equilibrium flow enables temperature-gradient-driven turbulence is identified. Gyrokinetic simulations of ion-scale plasma turbulence show that imposed {equilibrium \exb{} flow} shear can destroy the self-generated zonal flows that regulate the turbulence. {This results in} transport {that increases sharply with flow shear}. A reduced fluid model {demonstrates that this is due} to the {spatial} incompatibility of imposed and zonal shear layers. Simulations of spherical tokamak discharges place the inferred rotation shear at, or just below, the threshold of the sharp transport increase, implying that the toroidal rotation can be determined primarily by heat, rather than momentum, injection.

\end{abstract}

\maketitle

\section{Introduction}
\label{sec:intro}

The self-generation of large-scale shear flows is a common theme in fluid and plasma turbulence \cite{rhines75,terry00,srinivasan12,tobias13}, from Jupiter's banded zonal jets \cite{vasavada05} to the zonal flows regulating turbulence in tokamak plasmas \cite{diamond05}. Closely related to the inverse cascade and effective negative turbulent viscosity in two-dimensional turbulence \cite{kraichnan67,kraichnan76}, these shear flows can reorganize the turbulent state, suppress linear instabilities, modify transport, and alter the mechanisms of turbulent saturation. What happens when such intrinsic flows interact with an externally imposed flow? A logical expectation might be that the two reinforce each other. We show, however, that in magnetized plasma turbulence, the imposed flow can instead destroy the self-generated one, leading to a dramatic increase in turbulent fluctuations and transport.

In tokamak plasmas, ion-temperature-gradient (ITG) turbulence provides a particularly clean example of the interaction of self-generated and imposed flows. {For a limited range of temperature gradients above the linear instability threshold, } quasi-stationary, large-scale \footnote{In this work, we use the adjective ``large-scale'' to describe zonal flows whose radial correlation length is larger than that of the nonzonal fluctuations. We find no evidence that the zonal correlation length scales with the radial size of the flux-tube domain. Were it present, such a scaling would have suggested that the zonal flows are ``meso-scale'' and have a radial wavelength determined by a combination of the micro- (e.g., $\rhoi$) and macroscales (e.g., the equilibrium).} zonal \exb{} flows, driven nonlinearly by the Reynolds stress of the {fluctuations, are known to suppress the turbulence} and sustain a low-transport state {---} the Dimits regime \cite{DimitsShift2000} --- by shearing the ITG fluctuations. Additionally, these plasmas often carry a sheared equilibrium \exb{} flow, typically driven by external momentum injection from neutral-beam heating. The textbook picture {---} supported by extensive theory~\cite{Staebler_1991_linear,Waltz95_linearExBshear,Wang_ITGlinearstability,Biglari1990,Shaing_1990_POF_exbflowshear,zhang_1992_edge_scaling,NewtonFlowShearUnderstanding2010,HighcockRotationBifurcation2010,barnes2011,Roach_2009,ChristenFlowShear2018,Sun2024NF} and experiment~\cite{Ninomiya_1992_ExBrate_stablization_JT60,Kondoh_1994_ExBrate_destabilization_JT60,Lazarus_1996_ExBrate_stabilization_DIIID,JETdeVries_2006,JETdeVries_2008,Ida2009PRL,JETMantica2009PRL,Angioni2011IntrinsicRotation,schaffner12,ghim14,howard16} {---} posits that the externally imposed flow shear suppresses cross-field turbulent transport. 

In this Letter, we show that this expectation can fail in the {low-transport} Dimits regime. Using {local} nonlinear gradient-driven gyrokinetic {(GK)} simulations~\cite{Frieman_Chen_1982,Dubin_1983_Nonlineargyrokinetics,Hahm2007NLtheory,Brizard_2007_RMP,AbelGyrokineticsDeriv2012,catto19} of ITG turbulence, we demonstrate that the transport response to imposed flow shear is strongly non-monotonic. Weak imposed shear merely reorganizes the zonal-flow profile with little effect {on transport, consistent with the flux-driven study} of \cite{seiferling19}. Once the imposed shear becomes comparable to the intrinsic zonal shear, the zonal flows break down and the transport rises sharply before the turbulence is quenched at larger shear. We argue that this breakdown is unavoidable: the imposed shear prevents the Dimits-state pattern of alternating zonal-shear regions from forming and suppressing the turbulence.

A reduced two-dimensional fluid model of ITG turbulence reproduces the same behavior {as observed in tokamak GK simulations}, confirming that the mechanism does not depend on kinetic effects or toroidal geometry. We formalize the breakdown of the zonal flows as a geometric incompatibility between the imposed shear and the allowable widths of zonal shear regions. Since the underlying ingredients --- quasi-two-dimensional turbulence, Reynolds-stress drive of large-scale flows, and externally imposed shear --- are not specific to plasmas, this mechanism may {occur in} other systems with a mix of intrinsically driven and externally forced flows. 

Finally, we show that this effect is exacerbated in the newly identified tight-aspect-ratio, low-safety-factor regime of spherical tokamaks called the low-momentum-diffusivity regime, where the Dimits regime is particularly large. Gyrokinetic simulations of Mega Ampere Spherical Tokamak Upgrade (MAST-U) {discharges} place the experimentally inferred rotation shear at, or just below, the threshold for the breakdown of the zonal flows, suggesting that the toroidal rotation of spherical tokamaks can be governed by the interplay between zonal flows and external heat and momentum injection rather than by external momentum injection alone. These results overturn textbook expectations that equilibrium shear monotonically improves confinement and identify a generic interaction between externally imposed and self-organized shear in turbulent systems.

\section{Numerical results}
\subsection{Gyrokinetics}
\label{sec:gk_numerics}

We use the well-benchmarked code {\texttt{GENE}} \cite{JenkoGENE2000,GoerlerGENE2011} to perform local {(flux-tube)} linear and nonlinear electrostatic {GK} simulations of ITG-driven turbulence with adiabatic electrons \cite{hammett93}\footnote{{A limited set of kinetic-electron runs {follows} the behavior discussed in \cref{sec:gk_numerics}, but their cost precludes the parameter scans reported here.}}. Our reference case is a Miller equilibrium~\cite{Millergeometry1998,belli08,Snoep2023MillerGeneral} parametrization of a spherical tokamak with elongation $\kappa=1.5$, low safety factor $q=1.05$, tight aspect ratio $\epsilon=0.36$, and magnetic shear $\hat{s}=0.8$, representative of the low-momentum-diffusivity regime \cite{mcmillan2019,Sun_2025_NF} {(Table~S1 of {the Supplemental Material \cite{supp}})}.{\ While these idealized scans use adiabatic electrons, the MAST-U simulations of \cref{sec:MAST_experiment} retain kinetic electrons, electromagnetic fluctuations, and collisions.}

At long wavelengths {($\kperp\rhoi \ll 1$)}, equilibrium and zonal \exb{} flows enter the ion GK equation through {the} advection {term}
\begin{equation}
    (\vec{V} + \vec{V}_\text{ZF})\cdot\grad \avgRi{\dfi} = (V_\perp + V_\text{ZF}) \partd{\avgRi{\dfi}}{y},
    \label{eq:total_flow}
\end{equation}
where $\avgRiinline{\dfi}$ is the gyroaveraged perturbed ion distribution function, $y$ is the binormal coordinate, and $V_\perp = \vec{V}\cdot\grad y$ and $V_\text{ZF} = \vec{V}_\text{ZF}\cdot\grad y$ are the binormal projections of the mean and zonal \exb{} flow, respectively. Thus, the turbulent eddies experience the \emph{total} perpendicular shear, namely, the sum of \(\omega_\perp\equiv\partial_x V_\perp\)\footnote{A common alternative notation for \(\omega_\perp\) is \(\gamma_E\).} and \(\omega_{\rm ZF}\equiv\partial_x V_{\rm ZF}\), where $x$ is the radial coordinate. As tokamak equilibrium flows are toroidal, there is also {an associated} parallel-velocity-gradient (PVG) term, {which drives the PVG instability \cite{CattoPVG1973}}. Below, we show that PVG is not responsible for the behavior reported here.

\begin{figure}
  \centering
  \includegraphics[width=\columnwidth]{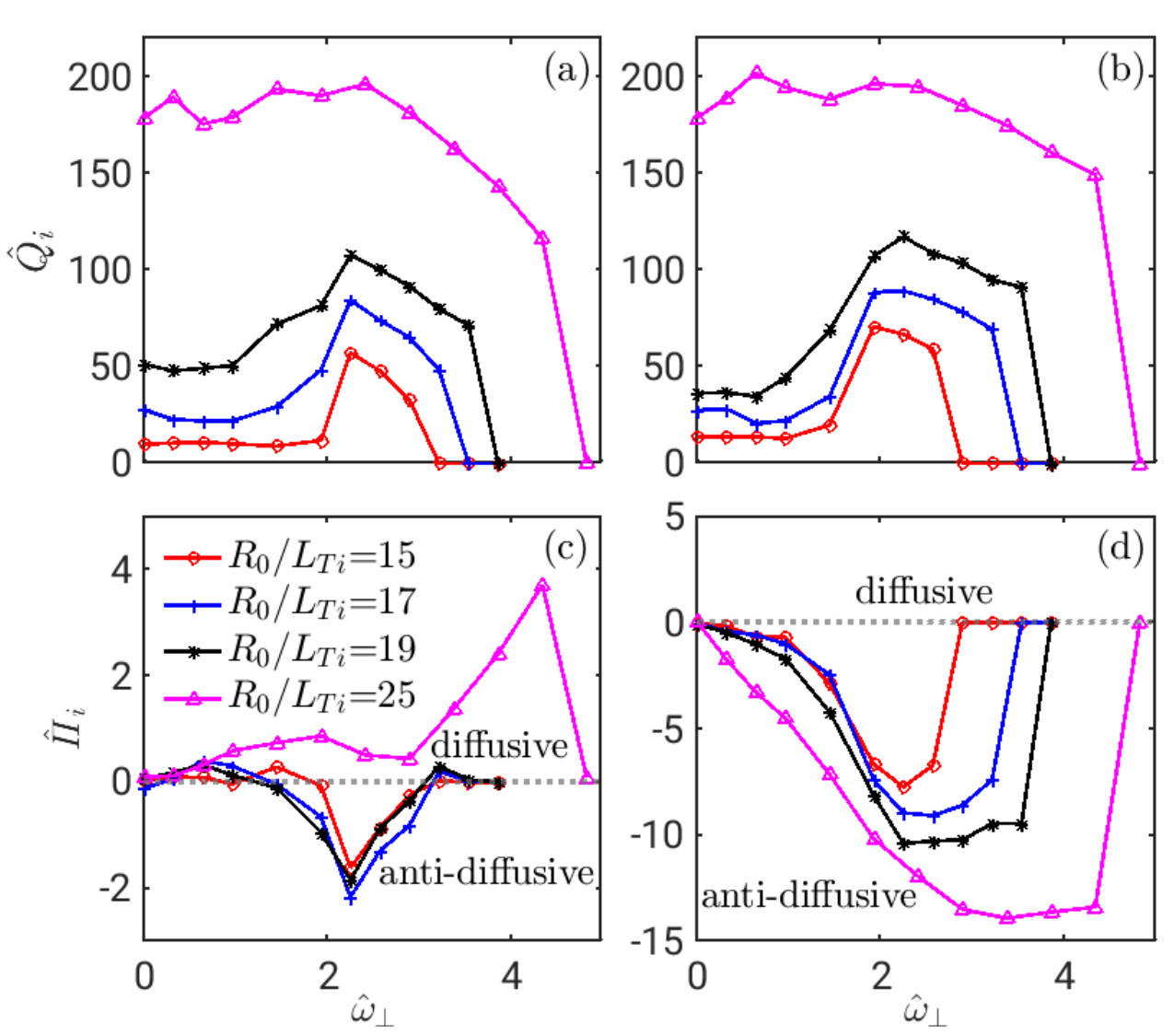}
  \caption{\texttt{GENE} simulations of the idealized Miller {equilibrium} (Table~S1 of the Supplemental Material \cite{supp}). {Normalized} ion heat flux $\hat{Q}_i$ (a,b) and ion toroidal angular momentum flux ${\hat{\varPi}_i}$ (c,d) versus \exb{} flow shear at several values of the inverse ITG length scale $R_0/\LTi$, where $R_0$ is the major radius at the center of the simulation domain. Only ${R_0/\LTi} = 25$ is above the Dimits threshold. Panels (a,c) use {full} toroidal flow shear; in (b,d) the PVG terms are artificially removed. The dotted gray lines in (c,d) mark ${\hat{\varPi}_i} = 0$, separating diffusive (${\hat{\varPi}_i} > 0$) from anti-diffusive (${\hat{\varPi}_i} < 0$) momentum transport. For normalizations, see the Supplemental Material \cite{supp}.}
  \label{fig:GK_scan1}
\end{figure}

\begin{figure}
  \centering
  \includegraphics[width=\columnwidth]{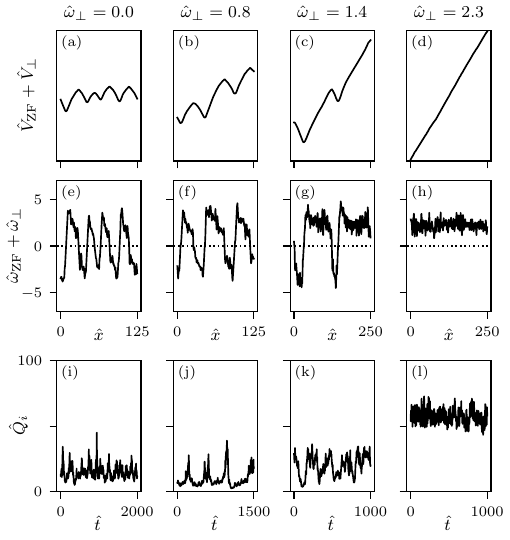}
  \caption{Time-averaged radial profiles of the total (zonal + mean) flow (a--d) and flow shear (e--h) and time traces of the normalized heat flux (i--l) obtained from tokamak \texttt{GENE} simulations of the idealized Miller equilibrium with $R_0 / \LTi = 15$ for four equilibrium flow shear values indicated above each column. {The $y$ axis of (a)--(d) is in arbitrary unlabeled units (same for the pairs of simulations with equal radial box size) as the absolute value of the total flow has no physical significance in local low-Mach-number GK.}}
  \label{fig:GK_spatialtmp}
\end{figure}

\Cref{fig:GK_scan1} shows the radial fluxes of ion heat and toroidal angular momentum as a function of the imposed \exb{} shear for several values of the ion temperature gradient. Below the Dimits threshold~\cite{DimitsShift2000} ({${R_0/\LTi} = 15\text{ (red)},17\text{ (blue)},\text{ and }19\text{ (black)}$} curves in \cref{fig:GK_scan1}), the {heat} transport is initially insensitive to \(\omega_\perp\), then \emph{increases} markedly as \(\omega_\perp\) approaches the characteristic zonal shear at $\omega_\perp = 0$ {given by $\omegac = \langle|\omega_\text{ZF}|\rangle_{\omega_\perp = 0} \approx 3 c_s / R_0$ } {(see \cref{fig:GK_spatialtmp}(e))}, and is suppressed only at larger shear. The same non-monotonic response is obtained when the PVG terms are removed (\cref{fig:GK_scan1}(b,d)), demonstrating that PVG is not {responsible for this {behavior}}. Henceforth, we refer to the intermediate-shear increase in the transport as the \emph{\exb{} destabilization} of the Dimits state.

The Dimits state is characterized by long-wavelength, slowly evolving (compared to the turbulent time scale) zonal flows that regulate the turbulence according to predator--prey dynamics. This produces bursty transport \cite{kobayashi12,kobayashi15,rath16,weikl17}, consistent with the first three columns of \cref{fig:GK_spatialtmp}. As \(\omega_\perp\) is increased (\cref{fig:GK_spatialtmp}, middle columns), the zonal-flow shear becomes asymmetric: broad regions of one sign{, where zonal and background shears add up,} alternate with narrower regions of the opposite sign{, where the shears oppose each other,} such that \(|\omega_\perp+\omega_{\rm ZF}|\) remains close to {the} critical value $\omegac$ over much of the domain. This is consistent with the Reynolds-stress drive arising from eddies sheared by the \emph{total} perpendicular flow as they cannot distinguish equilibrium from {long-wavelength} zonal shear, as seen in \cref{eq:total_flow}. {The flow shear locally breaks the up-down symmetry of the equilibrium and allows a nonzero perpendicular momentum flux \cite{ParraUpDownSym2011}. In the Dimits regime, this momentum flux builds up the zonal flows until the fluctuations are suppressed, which happens when the total local shear becomes comparable to $\omegac$.}

{Additionally, we have verified that the \exb{} destabilization is not related to turbulence bistability \cite{Christen_2022_exbrate_bistate}. Turning off the flow shear in the saturated state of $\hat{\omega}_\perp = 2.3$ of \cref{fig:GK_spatialtmp} returns the system to the low-transport Dimits state. Reintroducing the mean flow again at a later time breaks the zonal flows once again and recovers the state shown in the last column of \cref{fig:GK_spatialtmp}.}

At large enough $\omega_\perp$ (but still insufficient to quench the fluctuations alone), the Dimits-state zonal flows break down and the transport rises sharply. In the case shown in \cref{fig:GK_spatialtmp}, this results from the regions of negative zonal shear becoming too narrow to efficiently suppress the turbulence (see \cref{sec:cartoon}). As seen in \cref{fig:GK_scan1}(c), this is accompanied by a sign change of the angular momentum flux, corresponding to anti-diffusive momentum transport, i.e., the fluctuations are ``spinning up'' the plasma \footnote{To be more precise, the fluctuating contributions to the transport equations, which we do not solve here, would drive up the rotational shear at this flux surface.}.  {This suggests that, in contrast to the Dimits transition, where the effective zonal-flow turbulent viscosity changes from negative to positive \cite{ivanov20}, {the} breakdown of the zonal flows induced by the equilibrium rotation is not a consequence of a change in the response of the fluctuations to the flow shear.} {In the cases without PVG (\cref{fig:GK_scan1}(b,d)), the momentum transport is always anti-diffusive}, suggesting that parallel-momentum transport acts to damp the large-scale flows \cite{hallatschek2004_PRLzonal,seiferling18}.

As expected, larger flow shear is required to suppress the turbulence at larger temperature gradients (see {\cref{fig:GK_scan1}(a,b)}) as the instability growth rate increases with the gradient (see \cref{fig:linearrelation}). In contrast, the threshold for the \exb{} destabilization shifts to lower values of $\omega_\perp$ {at stronger ITG drive (\cref{fig:GK_scan1}).} This is consistent with larger gradients producing {radially} larger turbulent eddies (as expected from simple scaling theory \cite{barnes2011,Ivanov_2025_exbrate_stabilize,adkins26}), which then require {wider} regions of zonal shear to suppress the turbulence (see \cref{sec:cartoon}). 

Above the Dimits threshold {${{(R_0/\LTi)_\text{Dimits}\approx25}}$} (e.g., $R_0 / {\LTi} = 25$ in \cref{fig:GK_scan1}), large-scale zonal flows are absent, transport decreases monotonically with $\omega_\perp$, and momentum transport remains diffusive (when PVG terms are kept). \Cref{fig:dimits_scalings} (in End Matter) shows that {above the Dimits threshold}, the heat flux obeys the strongly driven scaling ${\hat{Q}_i \propto R_0/\LTi - (R_0/\LTi)_{crit}}${ \cite{Ghim_grand_critical,nies26,nies26a,adkins26}}, whereas below the transition it exhibits a much steeper dependence close to ${\hat{Q}_i} \propto (R_0 / {\LTi})^5$. This unusually large Dimits shift (compared to, e.g., the Cyclone base case {(CBC) }\cite{DimitsShift2000}) is characteristic of the low safety factor and tight aspect ratio \footnote{Here, $(q,\epsilon)=(1.05,0.36)$ versus $(1.4, 0.18)$ in CBC.} of the low-momentum-diffusivity regime (see \cref{fig:dimits_scalings}{(b)}) and is what makes \exb{} destabilization particularly prominent in this regime.

\subsection{Simplified fluid model}
\label{sec:fluid}

The shear-induced breakdown of the Dimits regime does not specifically depend on kinetic effects or toroidal geometry; rather, it follows from a small set of ingredients: ion-scale turbulence, Reynolds-stress generation of zonal flows, and externally imposed perpendicular shear. To isolate these ingredients, we study the two-dimensional cold-ion fluid model of ITG turbulence in a \(Z\)-pinch introduced in \cite{ivanov20}, augmented here by an imposed equilibrium flow shear. The governing equations and numerical methods are summarized in {the End Matter}.

\begin{figure}
  \centering
  \includegraphics[width=\columnwidth]{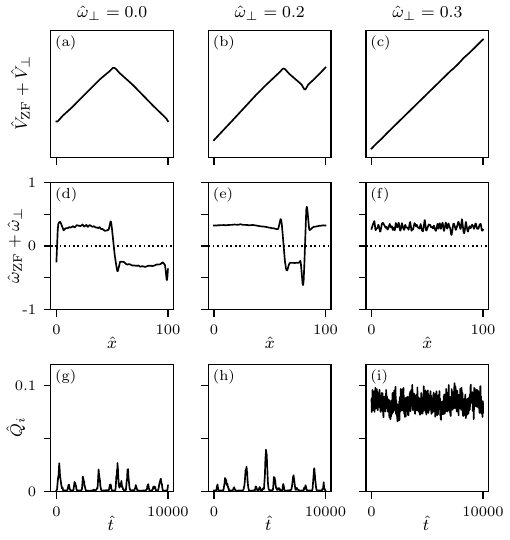}
  \caption{{{Same as \cref{fig:GK_spatialtmp} but for} the fluid model.}}
  \label{fig:fluid_zf_profiles}
\end{figure}

As in GK {simulations}, this model exhibits a Dimits regime {regulated by the long-wavelength, quasi-stationary ``triangular'' zonal flows of \cite{ivanov20,ivanov22}.} When \(\omega_\perp\) is small compared to the intrinsic zonal shear, the imposed mean shear reorganizes the zonal-flow profile so that the turbulence experiences nearly the same \emph{total} shear (zonal plus mean) as in the \(\omega_\perp=0\) case (\cref{fig:fluid_zf_profiles}(d,e)). In this regime, the heat flux is nearly independent of the mean flow shear. At larger \(\omega_\perp\), the system can no longer establish the Dimits-state zonal pattern: the zonal flows become {non-stationary and small-scale} and the heat flux increases by nearly an order of magnitude (\cref{fig:fluid_zf_profiles}(c,f,i)).

\iffalse
\begin{figure}
  \centering
  \includegraphics[width=\columnwidth]{figs/fluid_Q_Pi.pdf}
  \caption{stuff.}
  \label{fig:fluid_Q_Pi}
\end{figure}
\fi

\section{Dimits-incompatible rotation}
\label{sec:cartoon}

The \exb{} destabilization of the Dimits state might appear to be an artifact of local models, wherein the equilibrium shear is imposed and constant while the self-generated zonal flows are constrained by radial periodicity. However, we now show that there exists a general incompatibility between the Dimits-state zonal profiles {required for the low-transport state} and certain values of mean flow shear under the following assumptions: (i) the Dimits state requires the total local perpendicular flow shear (equilibrium plus zonal) to lie within a narrow range around a critical shear \(\omega_c\); (ii) each shear region must have a radial width of at least \(\ell_{\min}\) to suppress the turbulence; (iii) shear regions cannot exceed a maximum width \(\ell_{\max}\); (iv) the plasma's radial extent \(L\) is much larger than both \(\ell_{\min}\) and \(\ell_{\max}\).

\begin{figure}
    \centering
    \input{figs/cartoon.tikz}
    \caption{Illustration of the radial structure of the total \exb{} flow in the Dimits state with imposed mean flow shear.}
    \label{fig:cartoon}
\end{figure}
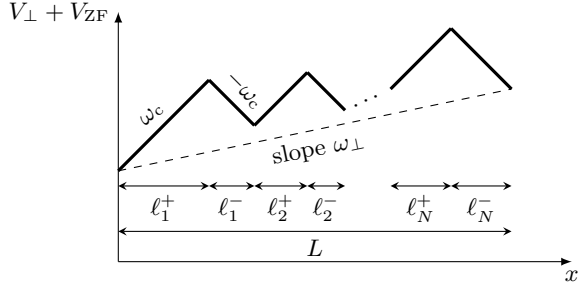

Let \(\omega_\perp\) denote the radially averaged equilibrium perpendicular flow shear over a domain of width \(L\). To maintain Dimits-regime turbulence suppression, the total \exb{} shear must organize into alternating regions of flow shear, as illustrated in \cref{fig:cartoon}. Here we idealize the local flow shear as equal to \(\pm\omegac\){,} and the overall flow difference adds up to $\omega_\perp L$. 

Thus,
{
    \begin{equation}
  \sum_{n=1}^N \ell^+_n\omegac - \sum_{n=1}^N \ell^-_n\omegac = \omega_\perp L.
  \label{eq:lnpm_omegaperp}
  \end{equation}
}
{Adding} and subtracting \cref{eq:lnpm_omegaperp} and the identity ${\omegac}\sum_n (\ell^+_n + \ell^-_n) = {\omegac}L$, we use the inequalities $\ell^+_n \leq \lmax$ and $\ell^-_n \geq \lmin$ to deduce that
\begin{equation}\label{eq:Ninequal}
    \frac{L}{\lmax}\left(1 + \frac{\omega_\perp}{\omegac}\right) \leq 2N \leq \frac{L}{\lmin}\left(1 - \frac{\omega_\perp}{\omegac}\right),
\end{equation}
{which can be rearranged into the compatibility condition:}
\begin{equation}
    \frac{\omega_\perp}{\omega_c} \le \lambda \equiv \frac{\ell_{\max}-\ell_{\min}}{\ell_{\max}+\ell_{\min}}.
    \label{eq:omega_perp_condition}
\end{equation}
{Assuming} \(\ell_{\min}>0\) and \(\ell_{\max}<L<\infty\) {then implies} \(\lambda<1\) and {hence} there exists a finite interval of instability{, \(\lambda < \omega_\perp/\omega_c < 1\),}
for which the imposed shear is too large to remain compatible with the Dimits-state zonal organization, yet too small to suppress the turbulence directly. The two {endpoints of this interval} have distinct physical origins: at $\omega_\perp = \lambda\omegac$, \cref{eq:Ninequal} first fails to admit an integer number of shear regions satisfying both width bounds, while at $\omega_\perp = \omegac$ the imposed shear alone reaches the quench scale and suppresses the turbulence without zonal-flow assistance. This is precisely the regime in which we observe transport increasing with imposed shear. {In that case, we find $\omegac \approx {2.5\text{--}3\,c_s/R_0}$, as evident from \cref{fig:GK_spatialtmp}(e,f,g), and the breakdown of zonal flows occurs at $\omega_\perp \approx {2\text{--}2.5\,c_s/R_0}$ with full turbulence suppression at $\omega_\perp \approx 3{\,c_s/R_0}$.} {The estimates of \(\ell_{\min}\) and \(\ell_{\max}\), and their comparison with simulations, are given in {the Supplemental Material \cite{supp}}.}

{Assuming a fixed number $N$ of shear regions, an increasing positive $\omega_\perp$ is accommodated by wider positive-shear and/or narrower negative-shear regions, consistent with \cref{eq:Ninequal} and \cref{fig:GK_spatialtmp,fig:fluid_zf_profiles}. If the width of the positive-shear regions is already at $\lmax$, then the shear regions break up and $N$ increases in order to support the increase in $\omega_\perp$. In contrast, when the negative-shear regions become limited by $\lmin$, $N$ decreases by }the merging of shear regions into wider ones. {This is qualitatively consistent with the widening regions of positive zonal shear seen in \cref{fig:GK_spatialtmp},} suggesting that{, in the cases shown there,} the system {sits close to} the $\ell^-_n\gtrsim\ell_{\min}$ boundary. 

Finally, note that the assumptions \(\ell_{\min}>0\) and \(\ell_{\max}<\infty\) are physically well motivated. A nonzero \(\ell_{\min}\) is expected because zonal flows narrower than individual turbulent eddies are inefficient at suppressing transport~\cite{hahm99,terry00}. In both GK and reduced-fluid simulations, we also observe a maximum zonal-flow radial scale that is independent of the radial box size (once the box is sufficiently large), implying a finite \(\ell_{\max}\) \cite{supp}. While the mechanism setting \(\ell_{\max}\) remains to be established, one plausible interpretation is that sustaining a shear region of a given width requires radial communication across the shear region. Both our GK and fluid simulations indicate that turbulence within the shear regions is dominated by propagating structures (``avalanches'' \cite{mcmillan09,villard13,seiferling18,seiferling19} or ``ferdinons'' \cite{wyk2016,wyk17,ivanov20}). {A} finite radial mean free path of these structures {could provide} a natural bound on \(\ell_{\max}\).

\section{Experimental results}\label{sec:MAST_experiment}

The mechanism outlined in this 
Letter should have observable consequences in tokamaks where external momentum drive is strong but the ion heat input is not sufficient to move the plasma across the heat-flux peak. In this case, the equilibrium shearing rate should be constrained to the left of the peaks in \cref{fig:GK_scan1}(a), i.e., {$\omega_\perp < \lambda \omegac$}. 
\begin{figure}
  \centering
  \includegraphics[width=1.05\columnwidth]{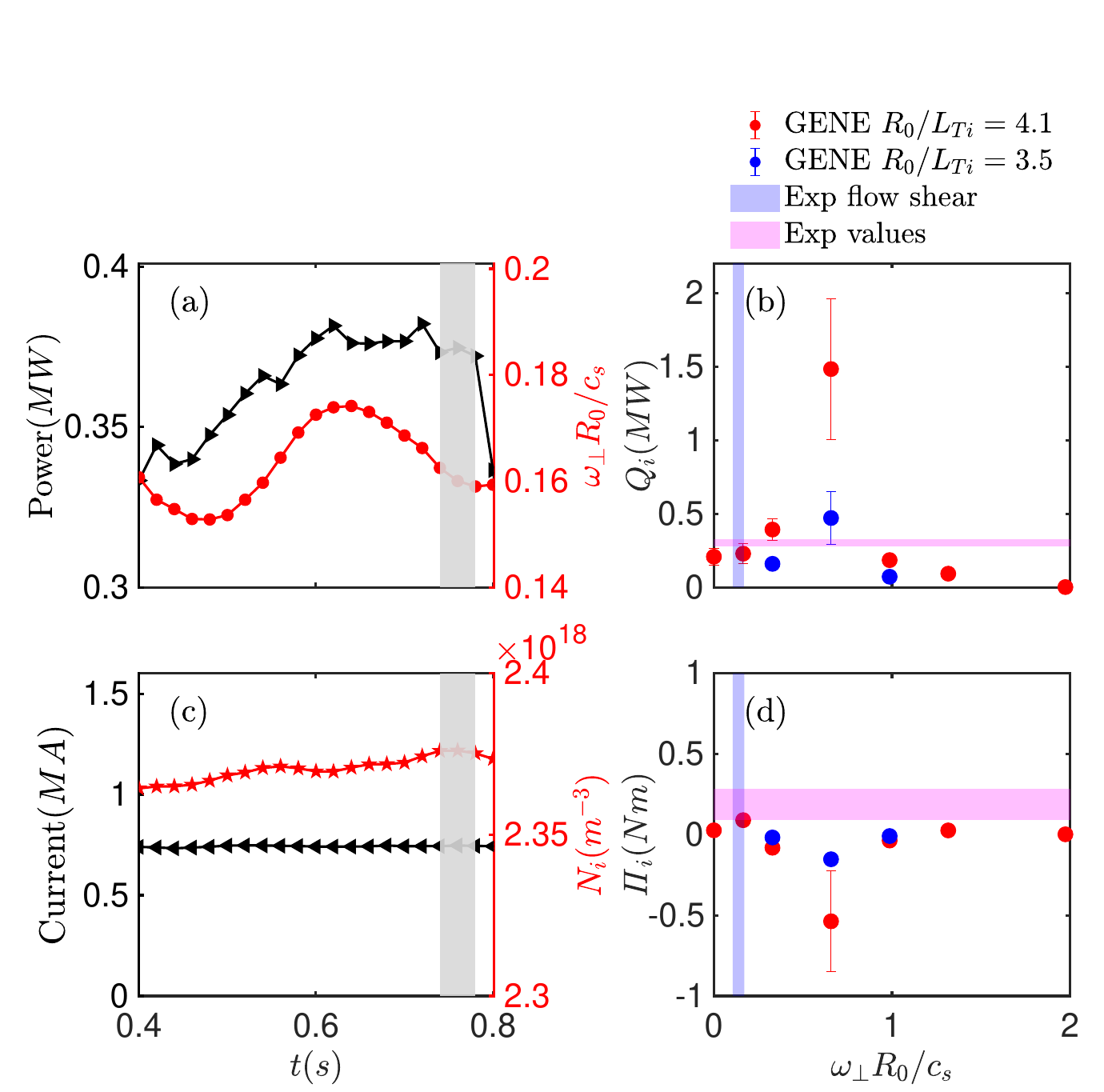}
  \caption{Time traces of {key experimental quantities} of MAST-U {shot 51653 (a,c), and (b) ion heat flux} and (d) ion toroidal angular {momentum flux versus flow shear at ${\rho_\text{tor}}=0.5$ and $t=0.77\,$s.} {The shaded regions in (a,c) denote the simulation time.} The red dots in (b,d) {{denote} the {\texttt{GENE} simulations} for experimental parameters. The blue dots are those with a lower ion temperature gradient.} The blue and magenta shaded regions in (b,d) show the experimental flow shear and fluxes, respectively, {with widths denoting the measurement {uncertainties}.}}
  \label{fig:MASTU}
\end{figure}

As an example, we consider shot 51653 of MAST-U at time $t=0.77\,$s and at radial location ${\rho_\text{tor}}=0.5${, where $\rho_\text{tor}$ is the square root of the normalized toroidal flux}. This is an L-mode plasma with on-axis {co-current neutral-beam injection} starting from $t=0.1\,$s, which drives differential rotation. {Detailed} experimental parameters are given in {Table~S2 of the Supplemental Material \cite{supp}}. \Cref{fig:MASTU}(a,c) {show that the plasma has reached a quasi-steady state in heating power, plasma current, and ion density; the rotation is stable, although its shear is still slowly evolving.} We perform a scan in the toroidal flow shear in local flux-tube \texttt{GENE} simulations in generalized Miller geometry{~\cite{Snoep2023MillerGeneral}} {with parameters fitted to this flux surface.} Kinetic electrons{,} electromagnetic fluctuations{, and collisions} are retained throughout. Linear simulations show that ITG is the dominant instability. \Cref{fig:MASTU}(b) shows that the simulated nonlinear heat flux 
rises sharply for $\omega_\perp R_0/c_s\gtrsim {0.3}$ and peaks near $\omega_\perp R_0/c_s\approx 0.66$; the destabilization onset lies just above the experimentally inferred rotation shear (blue shaded region). {This result is consistent with the experimental flow shear being limited by the \exb{} destabilization of the Dimits regime.} 
\Cref{fig:MASTU}(d) shows that the toroidal angular momentum flux becomes negative when the destabilization occurs, {as in} the idealized geometry. 
The simulated heat and momentum fluxes match the experimental ones reasonably well.
To test robustness, we decreased the ion temperature gradient within the experimental error bar (to $R_0/L_{Ti}=3.5$) and performed more nonlinear {GK} simulations. The blue solid dots in \cref{fig:MASTU} show that the destabilization persists \footnote{For MAST-U shot 51653, we also performed global, gradient-driven \texttt{GENE} simulations and observed the same \exb{} destabilization.}. The behavior above is also observed for other MAST-U shots, including but not limited to 44573, 44682, 46813, 51651, {and }52375 (see End Matter).

\section{Conclusions and discussion}

We have identified a regime in which externally imposed equilibrium flow shear increases, rather than suppresses, turbulent transport in magnetized plasmas. Nonlinear gyrokinetic simulations show that the effect arises from a breakdown of the Dimits state, in which self-organized zonal flows regulate the turbulence. The low safety factor and tight aspect ratio attainable in spherical tokamaks yield an unusually large Dimits shift, which exacerbates the transport jump associated with the breakdown and makes it more straightforward to identify. The effect persists when PVG terms are artificially removed, ruling out a PVG-instability interpretation{, and is found to be unrelated to turbulence bistability \cite{Christen_2022_exbrate_bistate}}. A simple geometric argument shows that the Dimits-regime breakdown is generic: {under the assumptions stated in \cref{sec:cartoon}, there exists an interval of values for the imposed flow shear that is incompatible with the self-organized Dimits state}. The same behavior is reproduced in a reduced cold-ion ITG fluid model, demonstrating that the mechanism is rooted in a few simple ingredients---ion-scale turbulence, Reynolds-stress drive of zonal flows, and externally imposed coherent shear---rather than in the details of GK or toroidal geometry. 

{A flow-shear-induced transport enhancement has been reported in multiscale electron-temperature-gradient turbulence \cite{belli24}, where it was associated with reduced ion-scale zonal energy and attributed to ion parallel dynamics and geodesic-acoustic modes. In contrast, the effect reported here is reproduced in a two-dimensional fluid model without these ingredients and is traced to the incompatibility of imposed and self-organized zonal shear. Whether the two phenomena share a broader connection is an open question.
}

The MAST-U analysis points to a qualitatively new operating constraint. In the standard picture, in which equilibrium \exb{} shear monotonically suppresses transport \cite{casson2009,Angioni2011IntrinsicRotation}, the steady-state rotation shear is set by momentum diffusivity alone. The mechanism reported here introduces a heat-flux ``hill'' between the low-transport Dimits state and the strongly suppressed regime. For the rotation to ``climb'' past the ``hill'', the device must inject enough heat to sustain {the plasma profiles} against the enhanced {transport} by {the \exb{} flow shear destabilization}. The achievable rotation shear is therefore limited mainly by the heat injection.%

Beyond plasma physics, the mechanism identifies a new mode of interaction between externally imposed shear and self-organized zonal shear in a turbulent system: when the two are of comparable strength, the geometric constraints on the self-organized shear pattern can no longer be reconciled with the constraints set by the imposed shear, and the regulating {self-organized} structure breaks down. Because the underlying ingredients are not specific to plasmas, the same scenario may arise in other turbulent systems, such as atmospheric and oceanic zonal jets driven against an externally imposed mean shear \cite{Panetta1993JAS,FarrellIoannou2008JAS,Berloff2009JFM}.

\textbf{\textit{Acknowledgments{---}}}The simulations in this work {were} performed on CSCS Daint and Cineca Pitagora. The authors thank Prof. Per Helander, Dr. Josefine Proll, Dr. Eduardo Rodriguez, Dr. Francis Casson, Dr. Arnas Volcokas{,} and Dr. Alessandro Balestri for {fruitful discussions}. This work was supported by a grant from the Swiss National Supercomputing Centre (CSCS) under project {IDs} lp34 and lp134. This work has been carried out within the framework of the EUROfusion Consortium, partially funded by the European Union via the Euratom Research and Training Programme (Grant Agreement No. 101052200 - EUROfusion). The Swiss contribution to this work has been funded by the Swiss State Secretariat for Education, Research and Innovation (SERI). Views and opinions expressed are however those of the author(s) only and do not necessarily reflect those of the European Union, the European Commission or SERI. Neither the European Union nor the European Commission nor SERI can be held responsible for them. This work was supported in part by the Swiss National Science Foundation and by the EPSRC Energy Programme (Grant Number EP/W006839/1). Part of this work was performed using resources provided by the Cambridge Service for Data Driven Discovery (CSD3) operated by the University of Cambridge Research Computing Service (\texttt{www.csd3.cam.ac.uk}), provided by Dell EMC and Intel using Tier-2 funding from the Engineering and Physical Sciences Research Council (capital grant EP/T022159/1), and DiRAC funding from the Science and Technology Facilities Council (\texttt{www.dirac.ac.uk}). 

\makeatletter
\immediate\write\@auxout{\string\citation{apsrev42Control}}
\makeatother
\bibliographystyle{apsrev4-2} %
\bibliography{apssamp}%

\clearpage
\section*{End Matter}

\textit{Fluid model---}The equations of the fluid model considered in \cref{sec:fluid} are
\begin{align}
    &\der{}{t}\left(\nonzonal{\phinorm} - \rhosound^2 \gradperp^2\phinorm \right) + \omega_\perp\rhosound^2\frac{\partial^2T}{\partial x \partial y} - \frac{\rhoi\vthi}{\tau\LB}\partd{}{y}\left(\phinorm + T\right) \nonumber \\ &\quad+ \frac{\rhoi\vthi}{2\LTi}\partd{}{y}\left(\rhosound^2\gradperp^2\phinorm\right)  + \frac{1}{2}\rhoi\vthi\rhosound^2\gradperp \cdot \pbra{\gradperp \phinorm}{T} \nonumber \\ &\quad= -\chi \rhosound^2\gradperp^4 \left(\frac{9}{40}\phinorm - \frac{67}{160}T\right),
    \label{eq:fluidphieq} \\
    &\der{T}{t} + \frac{\rhoi\vthi}{2\LTi} \partd{\phinorm}{y} = \chi\gradperp^2 T{,} \label{eq:fluidTeq}
\end{align}
where $\LTi^{-1} \equiv {-}\partial_x \ln T_i$ is the ion temperature gradient, $\LB^{-1} \equiv {-}\partial_x \ln B$ is the gradient length scale of the magnetic field, $\phinorm \equiv Z_i e \phi / \Ti$ is the normalized electrostatic potential{\ ($Z_i$ is the ion charge number, $e$ the elementary charge, $\phi$ the electrostatic potential, and $\Ti$ the ion temperature)}, $T = \dTi / \Ti$ is the normalized perturbed ion temperature, $\tau = \Ti / {Z_i}\Te \ll 1$ is the temperature ratio{\ ($\Te$ is the electron temperature)}, $\rhosound \equiv \rhoi / \sqrt{2\tau}$ is the sound radius{, where $\rhoi = \vthi/\Omegai$ is the ion gyroradius, $\vthi = \sqrt{2\Ti/\mi}$ the ion thermal speed, $\Omegai$ the ion cyclotron frequency, and $\mi$ the ion mass}, $\chi$ is the collisional thermal diffusivity, $\pbra{f}{g} \equiv {\ub} \cdot (\grad f \times \grad g)$ is the usual Poisson bracket{\ (with $\ub = \vB/B$ the unit vector along the magnetic field)}, and the $\derinline{}{t}$ operator includes advection by both the mean flow and the fluctuating \exb{} flows:
\begin{equation}
    \der{}{t} = \partd{}{t} + \omega_\perp x \partd{}{y} + \frac{1}{2}\rhoi\vthi (\ub \times \grad \phinorm)\cdot\grad.
\end{equation}
In \cref{eq:fluidphieq}, $\nonzonal{\phinorm} = \phinorm - \zonal{\phinorm}$ denotes the nonzonal component of the potential, where $\zonal{\phinorm}(x) \equiv \int (\rmd y / L_y) \phinorm(x, y)$ is the zonal average of $\phinorm$ and $L_y$ is the binormal size of the integration domain. \Cref{eq:fluidphieq,eq:fluidTeq} follow from the collisional, cold-ion, long-wavelength limit discussed in \cite{ivanov20}, where the reader can find a detailed derivation in the case of $\omega_\perp = 0$. The additional terms that arise in the presence of mean flow shear {are derived in Supplemental Material \cite{supp}, following} the arguments in Appendix A of \cite{ivanov20}.

For the results presented in \cref{sec:fluid}, \cref{eq:fluidphieq,eq:fluidTeq} are integrated using a pseudo-spectral, implicit-explicit method detailed in \cite{ivanov20, Ivanov_2025_exbrate_stabilize}. All simulations shown here have a resolution of $341\times341$ dealiased Fourier modes, box size $L_x = L_y = 100\rhosound$, normalized temperature gradient $\tau \LB / 2\LTi = 0.36$, and normalized thermal diffusivity $\chi\LB / 2\rhosound^3\Omegai = 0.1$. 

In \cref{fig:fluid_zf_profiles}, the normalized mean flow shear is $\hat{\omega}_\perp\equiv \omega_\perp \LB / 2\rhosound \Omegai$, while the normalized zonal flow and shear are $\hat{V}_\text{ZF}\equiv (\LB / 2\rhosound^2\Omegai) \partial_x \phi / B$ and $\hat{\omega}_\text{ZF} \equiv \rhosound \partial_x \hat{V}_\text{ZF}$.{\ The normalized radial coordinate is $\hat{x}\equiv x/\rhosound$ and the normalized mean flow is $\hat{V}_\perp \equiv \hat{\omega}_\perp \hat{x}$.} The normalized time is $\hat{t}\equiv 2\rhosound\Omegai t / \LB$. The heat flux is normalized as \(\hat{Q}_i = Q_i / Q_\text{norm}\), where \(Q_\text{norm} = 6n_iT_i{c_s} (\rhosound / \LB)^2 / \tau\){, with $c_s \equiv \rhosound\Omegai$ and $n_i$ the ion density}.

\iffalse
\begin{figure}
  \centering
  \includegraphics[width=\columnwidth]{figs/fluid_ferdinons.pdf}
  \caption{Snapshots of the instantaneous temperature fluctuations in the saturated state, normalized to the instantaneous maximum of $\absinline{T}$.}
  \label{fig:fluid_ferdinons}
\end{figure}
\fi

\iffalse
\section{Derivation of \cref{eq:omega_perp_condition}}
\label{appendix:cartoon}

For a zonal-flow profile as shown in \cref{fig:cartoon}, where we have idealized the local flow shear in the Dimits regime as equal to \(\pm\omegac\), the overall width and flow difference add up to $L$ and $\omega_\perp L$, respectively, we deduce that
\begin{equation}
    \sum_{n=1}^N \frac{\ell^+_n}{L} - \sum_{n=1}^N \frac{\ell^-_n}{L} = \frac{\omega_\perp}{\omegac}.
\end{equation}
Using $\ell^+_n \leq \lmax$ and $\ell^-_n \geq \lmin$, we can deduce that
\begin{equation}
    \frac{L}{\lmax}\left(1 + \frac{\omega_\perp}{\omegac}\right) \leq 2N \leq \frac{L}{\lmin}\left(1 - \frac{\omega_\perp}{\omegac}\right),
\end{equation}
which can then be rearranged into \cref{eq:omega_perp_condition}.
\fi
\vspace{1em}

\begin{figure}
  \centering
  \includegraphics[width=0.5\columnwidth]{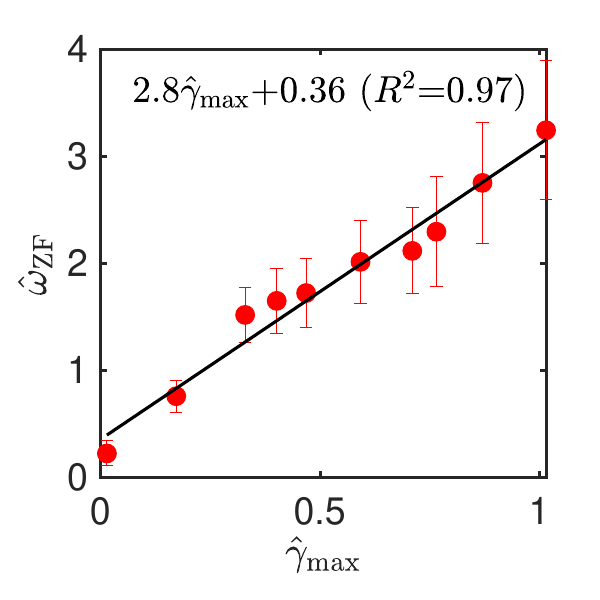}
  \caption{{Dependence of the average {zonal-flow} shear $\hat{\omega}_\text{ZF}$ in the zonal shear regions on the maximum linear growth rate $\hat{\gamma}_\text{max}$ for the {idealized Miller} geometry considered in \cref{sec:gk_numerics} {and described in {Table~S1 of the Supplemental Material \cite{supp}}}. The growth rate is varied by changing {${R_0/\LTi}$}.} All values of the temperature gradient are within the Dimits regime, i.e., ${R_0/\LTi < R_0/L_{Ti,\text{Dimits}}}\approx 25$ (see also \cref{fig:dimits_scalings}). The time averages are taken after the {zonal-flow} evolution has reached a steady state.}
  \label{fig:linearrelation}
\end{figure}

\begin{figure}
  \centering
  \includegraphics[width=\columnwidth]{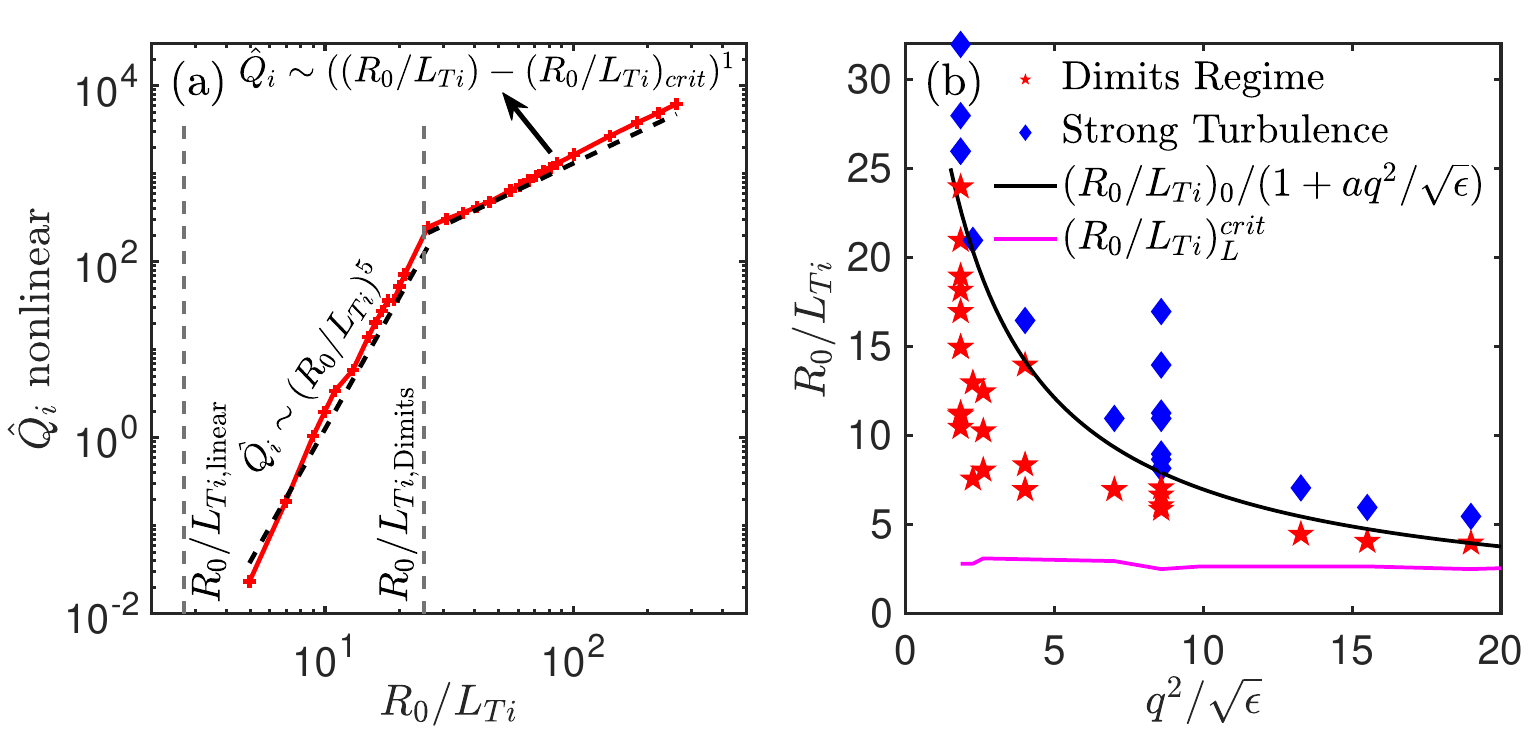}
  \caption{
  (a) Nonlinear heat flux $\hat{Q}_i$ (red, logarithmic $y$ axis) as a function of the ion temperature gradient $R_0/{\LTi}$ at zero flow shear for $q^2/\sqrt{\epsilon}=1.84$. The scaling properties of $\hat{Q}_i$ versus $R_0/{\LTi}$ are denoted by the black dashed lines, where {the scaling} changes from $\hat{Q}_i\sim (R_0/L_{Ti})^5$ to $\hat{Q}_i\sim ((R_0/L_{Ti})-(R_0/L_{Ti})_{{crit}})^1$ at $R_0/{\LTi} \approx 25$. {The vertical dashed gray lines denote the two critical gradients: the linear instability threshold ${R_0/L_{Ti,\text{linear}}}\approx 2.7$ [given by the magenta curve in (b) at $q^2/\sqrt{\epsilon}=1.84$] and the Dimits threshold ${R_0/L_{Ti,\text{Dimits}}}\approx 25$.} Another version of the plot in linear axes, showing the ultra-wide Dimits regime{,} is {shown in \cref{fig:wideDimits}}. (b) Dimits shift as a function of $q^2/\sqrt{\epsilon}$. The red stars denote the Dimits regime while the blue diamonds denote the strongly turbulent regime. The black line shows a parametric fit of the boundary of the two regimes{, of the form $(R_0/L_{Ti})_\text{Dimits} = (R_0/L_{Ti})_{0}/(1+a\,q^2/\sqrt{\epsilon})$}, where $(R_0/L_{Ti})_{0}=46$, $a=0.56$. The magenta line shows the linear critical temperature gradient $(R_0/L_{Ti})_{L}^{crit}$. The cases shown on this plot have {$q\in\{1,1.5,2,2.5,3\}$, $\epsilon\in\{0.18,0.24,0.36\}$, and $\hat{s}\in\{0.4,0.8,1.2,1.6\}$}. }
  \label{fig:dimits_scalings}
\end{figure}

\textit{Dimits regime of the low-momentum-diffusivity regime---}The low-momentum-diffusivity regime \cite{Sun_2025_NF}, characterized by a low safety factor and tight aspect ratio, exhibits a significantly larger Dimits shift than the standard {CBC} \cite{DimitsShift2000}. \Cref{fig:dimits_scalings}{(a)} shows the dependence of the ion heat flux {on} the ion temperature gradient (with no mean flow shear). Below the transition at $R_0/\LTi \approx 25$, we find a typical Dimits state dominated by alternating regions of zonal shear and a particularly steep scaling ${\hat{Q}_i} \propto (R_0/\LTi)^5$. \Cref{fig:linearrelation} shows that the average zonal shear in the Dimits state scales linearly with the maximum ITG growth rate (the ``quench'' rule \cite{Waltz1998GyroFluid,waltz94}). The strongly turbulent state at $R_0/\LTi > 25$ exhibits a linear scaling of the heat flux as predicted by the theory of ``grand critical balance'' \cite{Ghim_grand_critical,nies26,nies26a,adkins26}.

Furthermore, we have found that the Dimits state is most sensitive to the safety factor and aspect ratio. In \cref{fig:dimits_scalings}{(b)}, we plot the critical gradient for the Dimits transition as a function of ${q^2/\sqrt{\epsilon}}$ for a set of simulations with varying $q$, $\epsilon$, and $\hat{s}$. We find an empirical scaling $R_0/\LTi \propto 1/(1+aq^2/\sqrt{\epsilon})$. While we do not have a satisfactory theory that predicts or explains this dependence, the {presence} of $\sqrt{\epsilon}$ suggests that trapped particles and orbit widths are {important ingredients} in the Dimits transition. The appearance of such effects is not entirely surprising as they are also responsible for the scaling of the Rosenbluth{--}Hinton residual {\cite{RosenbluthZonalFlowDamping1998} }with $1/(1 + 1.6q^2/\sqrt{\epsilon})$ in the $\epsilon \ll 1$ limit. {A first-principles explanation of the Dimits transition and its dependence on toroidal geometry is left for future work}.

\begin{figure}
  \centering
  \includegraphics[width=0.65\columnwidth]{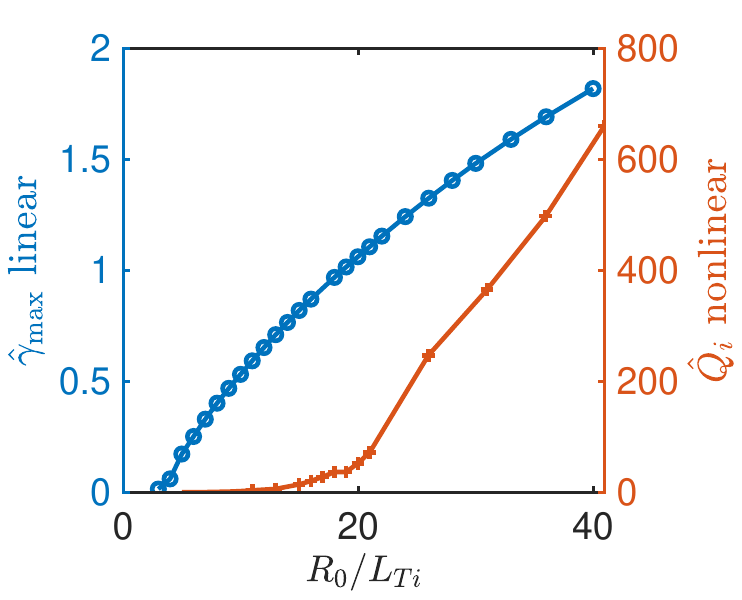}
  \caption{Maximum linear growth rate (left axis) and nonlinear heat flux (right axis) as a function of $R_0/L_{Ti}$ in idealized {Miller} geometry (same as \cref{fig:linearrelation}).}
  \label{fig:wideDimits}
\end{figure}

\textit{Ultra-wide Dimits regime---}\Cref{fig:wideDimits} shows the ultra-wide Dimits regime that we see in our cases without external flow shear. The left axis of the figure shows the maximum linear growth rates as a function of $R_0/L_{Ti}$, whereas the right axis of this figure {shows} another version of \cref{fig:dimits_scalings}(a) in linear axes.

\textit{{\exb{}} destabilization in other MAST-U discharges---}{The destabilization reported in \cref{sec:MAST_experiment} is not unique to shot 51653. We have performed nonlinear GK simulations of five additional MAST-U discharges: 44573, 44682, 46813, 51651, and 52375. For each discharge, we fit a generalized Miller equilibrium to a flux surface in the outer core region ($\rho_{\text{tor}} \approx 0.5$--$0.6$) and scan the imposed toroidal flow shear, keeping all other parameters fixed. {The ITG instability is the dominant instability and source of fluctuations} for all the MAST-U cases considered in this Letter. As in \cref{sec:MAST_experiment}, kinetic electrons, electromagnetic fluctuations, and collisions are retained, and the imposed flow shear is purely toroidal.
}

\begin{figure}[h]
  \centering
  \includegraphics[width=0.8\columnwidth]{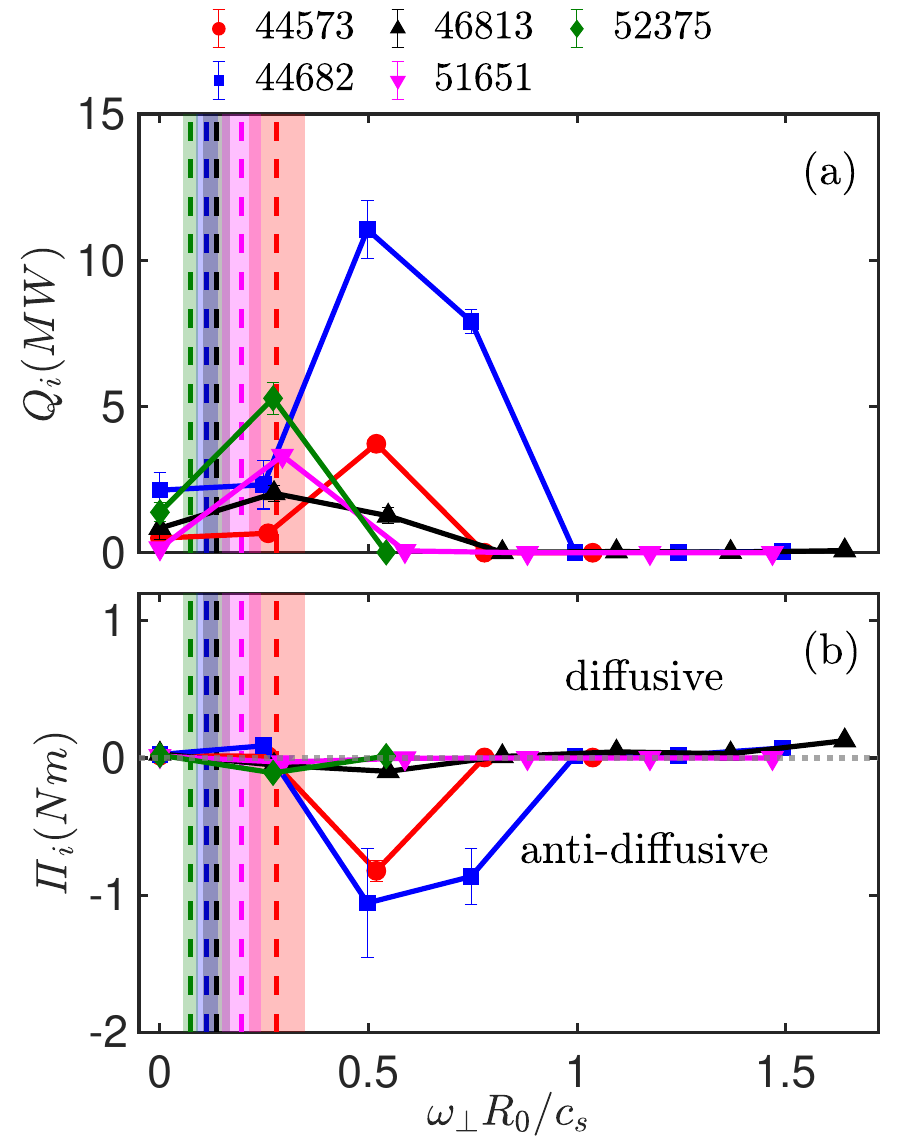}
  \caption{{(a) Ion heat flux {$Q_i$ (in $MW$) }and (b) ion toroidal angular momentum flux {$\varPi_i$ (in $Nm$) }as functions of the imposed flow shear from nonlinear \texttt{GENE} simulations of five additional MAST-U discharges (shot numbers in the legend). The dotted gray line in (b) marks ${\varPi_i} = 0$, separating diffusive (${\varPi_i} > 0$) from anti-diffusive (${\varPi_i} < 0$) momentum transport. The dashed vertical lines mark the experimentally inferred rotation-shear magnitude of each discharge, in the same color as the corresponding curve. The shaded band of the same color shows the associated uncertainty, taken to be the same relative uncertainty as estimated for shot 51653 in \cref{fig:MASTU}(b). 
  }}
  \label{fig:othershots}
\end{figure}

{\Cref{fig:othershots} shows the time-averaged ion heat flux ${Q_i}$ and ion toroidal angular momentum flux ${\varPi_i}$ as functions of the imposed flow shear. All five shots exhibit the same behavior as shot 51653 in \cref{fig:MASTU}: the heat flux increases with $\omega_\perp$ before the turbulence is quenched at larger flow shear. As in \cref{fig:GK_scan1,fig:MASTU}, the destabilization is accompanied by an anti-diffusive toroidal angular momentum flux [\cref{fig:othershots}(b)], which is most prominent for shots 44573 and 44682.
 {\ For every discharge, the experimentally inferred rotation shear (vertical shaded bands in \cref{fig:othershots}) lies at, or below, the onset of the destabilization, as for shot 51653 in \cref{sec:MAST_experiment}. This is consistent with the differential rotation being limited by the associated increase in the heat transport.}}

\end{document}

%% file: cleveref_setup.tex
\usepackage{cleveref}

\crefformat{section}{Section~#2#1#3}
\Crefformat{section}{Section~#2#1#3}

\crefformat{subsection}{Section~#2#1#3}
\Crefformat{subsection}{Section~#2#1#3}

\crefformat{subsubsection}{Section~#2#1#3}
\Crefformat{subsubsection}{Section~#2#1#3}

\crefformat{appendix}{Appendix~#2#1#3}
\Crefformat{appendix}{Appendix~#2#1#3}

\crefformat{subappendix}{Appendix~#2#1#3}
\Crefformat{subappendix}{Appendix~#2#1#3}

\crefformat{subsubappendix}{Appendix~#2#1#3}
\Crefformat{subsubappendix}{Appendix~#2#1#3}

\crefformat{figure}{Figure~#2#1#3}
\Crefformat{figure}{Figure~#2#1#3}

\crefrangeformat{figure}{Figures #3#1#4--#5#2#6}
\crefmultiformat{figure}{Figures~#2#1#3}{ and~#2#1#3}{, #2#1#3}{, and~#2#1#3}

\Crefrangeformat{figure}{Figures #3#1#4--#5#2#6}
\Crefmultiformat{figure}{Figures~#2#1#3}{ and~#2#1#3}{, #2#1#3}{, and~#2#1#3}

\crefformat{table}{Table~#2#1#3}
\Crefformat{table}{Table~#2#1#3}
\crefrangeformat{table}{Tables #3#1#4--#5#2#6}
\crefmultiformat{table}{Tables~#2#1#3}{ and~#2#1#3}{, #2#1#3}{, and~#2#1#3}
\Crefmultiformat{table}{Tables~#2#1#3}{ and~#2#1#3}{, #2#1#3}{, and~#2#1#3}

\crefformat{equation}{(#2#1#3)}
\Crefformat{equation}{Equation~(#2#1#3)}

\crefrangeformat{equation}{(#3#1#4)--(#5#2#6)}
\crefmultiformat{equation}{(#2#1#3)}{ and~(#2#1#3)}{, (#2#1#3)}{, and~(#2#1#3)}

\Crefrangeformat{equation}{Equations~(#3#1#4)--(#5#2#6)}
\Crefmultiformat{equation}{Equations~(#2#1#3)}{ and~(#2#1#3)}{, (#2#1#3)}{, and~(#2#1#3)}

%% file: macros.tex
\makeatletter
\newcommand{\showfontsize}{%
  \f@size pt%
}
\makeatother
\def\rmd{{\rm d}}

\newcommand{\der}[3][\relax]{\frac{\rmd^{#1} #2}{\rmd {#3}^{#1}}}
\newcommand{\derinline}[3][\relax]{\rmd^{#1} #2/\rmd {#3}^{#1}}

\newcommand{\absinline}[1]{\rvert#1\rvert}

\newcommand{\zonal}[1]{{\overline{#1}}}
\newcommand{\nonzonal}[1]{{#1'}}

\newcommand{\grad}{{\vec{\nabla}}}
\newcommand{\gradperp}{\grad_\perp}
\ifdefined\partd
	\renewcommand{\partd}[3][]{\frac{{\partial^{#1} #2}}{{\partial #3}^{#1}}}
\else
	\newcommand{\partd}[3][]{\frac{{\partial^{#1} #2}}{{\partial #3}^{#1}}}
\fi

\newcommand{\uvec}[1]{{\hat{#1}}}

\ifdefined\vv
\renewcommand{\vv}{{\vec{v}}}
\else
\newcommand{\vv}{{\vec{v}}}
\fi

\newcommand{\kperp}{k_\perp}

\newcommand{\vths}[1][\s]{v_{\text{th}{#1}}}
\newcommand{\vthi}{\vths[i]}

\newcommand{\rhos}[1][\s]{\rho_{#1}}
\newcommand{\rhoi}{\rhos[i]}

\newcommand{\rhosound}{\rhos[\text{s}]}

\newcommand{\mss}[1][\s]{m_{#1}}

\newcommand{\mi}{\mss[i]}

\newcommand{\Ts}[1][\s]{T_{#1}}
\newcommand{\Te}{\Ts[e]}
\newcommand{\Ti}{\Ts[i]}

\newcommand{\dTs}[1][\s]{\delta T_{#1}}

\newcommand{\dTi}{\dTs[i]}

\newcommand{\dfs}[1][\s]{\delta f_{#1}}

\newcommand{\dfi}{\dfs[i]}

\newcommand{\pbra}[2]{\left\lbrace #1, #2 \right\rbrace}

\newcommand{\vRi}{{\vec{R}_i}}

\newcommand{\phinorm}{\varphi}  % normalised electrostatic potential

\newcommand{\vE}{\vec{E}}
\newcommand{\vB}{\vec{B}}
\newcommand{\ub}{\uvec{b}}

\newcommand{\exb}{\(\vE\times\vB\)}

\newcommand{\avgRi}[1]{\left\langle #1 \right\rangle_{\vRi}}

\newcommand{\avgRiinline}[1]{\langle #1 \rangle_{\vRi}}

\newcommand{\LB}{L_B}

\newcommand{\LTi}{\LT[i]}

\newcommand{\Omegas}[1][\s]{\Omega_{#1}}

\newcommand{\Omegai}{\Omegas[i]}

%% file: figs/cartoon.tikz
\begin{tikzpicture}
	\coordinate (origin) at (0,0);    
	
	% horizontal axis
	\def\xaxislength{6}
	\coordinate (xaxis) at ($ (\xaxislength,0,0) $);
	\draw[-latex] (0,0) -- (xaxis);
	\draw (xaxis) node[anchor=north] {$x$};
	
	% vertical axis
	\def\yaxislength{3.3}
	\coordinate (yaxis) at ($ (0, \yaxislength,0) $);
	\draw[-latex] (0,0) -- (yaxis);
	\draw (yaxis) node[anchor=east] {$V_\perp + V_\text{ZF}$};
	
	% --- Input arrays ---
	\def\A{1.2,0.7,0.8}   % positive slope (+1)
	\def\B{{0.6,0.5,0.8}}     % negative slope (-1)
	
	% --- Initial point ---
	\def\yoffset{1.2}
	\coordinate (start) at ($(origin) + (0, \yoffset)$);
	\coordinate (P) at (start);
	
	% --- width labels location ---
	\def\ylabels{1}
	\newlength{\Px}
%	\newdimen\Qx
	
	% --- Draw segments ---
	\foreach \a [count=\i] in \A {
		% +1 slope segment
		\coordinate (Q) at ($(P)+(\a,\a)$);
		
		\ifnum\i=1
			\draw[very thick] (P) -- (Q) node[midway,sloped,above] {$\omega_\text{c}$};
		\else
			\draw[very thick] (P) -- (Q);
		\fi
	
		\ifnum\i<3
			\draw[stealth-stealth] ($(P |- 0,\ylabels)$) -- ($(Q |- 0,\ylabels)$) node[midway,below] {$\ell^+_{\i}$};
		\else
			\draw[stealth-stealth] ($(P |- 0,\ylabels)$) -- ($(Q |- 0,\ylabels)$) node[midway,below] {$\ell^+_{N}$};
		\fi
		
		\coordinate (P) at (Q);
		
		% -1 slope segment
		\pgfmathparse{\B[\i-1]}
		\let\b\pgfmathresult
		\coordinate (Q) at ($(P)+(\b,-\b)$);
	
		\ifnum\i=1
			\draw[very thick] (P) -- (Q) node[midway,sloped,above] {$-\omega_\text{c}$};
		\else
			\draw[very thick] (P) -- (Q);
		\fi

		\ifnum\i<3
			\draw[stealth-stealth] ($(P |- 0,\ylabels)$) -- ($(Q |- 0,\ylabels)$) node[midway,below] {$\ell^-_{\i}$};
		\else
			\draw[stealth-stealth] ($(P |- 0,\ylabels)$) -- ($(Q |- 0,\ylabels)$) node[midway,below] {$\ell^-_{N}$};
		\fi
		
		\coordinate (P) at (Q);
		
		\ifnum\i=2
			\node (dots) at (P) [anchor=west,rotate=25] {$\dots$};
			\coordinate (P) at (dots.east);
		\fi

	}
	
	\draw[dashed] (start) -- (P) node[midway,below,sloped] {slope $\omega_\perp$};
	
	\def\yL{0.4}
	
	\draw[stealth-stealth] ($(start |- 0,\yL)$) -- ($(P |- 0,\yL)$) node[midway,below] {$L$};

\end{tikzpicture}